\documentclass[letterpaper, 12pt]{llncs}
\usepackage{amssymb}
\usepackage{graphicx}
\usepackage{subfigure}
\usepackage{comment}
\usepackage{color, soul}

\usepackage{url}
\urldef{\mailsa}\path|{stefan.goeller, alfred.hofmann, peter.strasser, lnicst}@springer.com|

\newcommand{\keywords}[1]{\par\addvspace\baselineskip
\noindent\keywordname\enspace\ignorespaces#1}

\newcommand{\bigO}{\ensuremath{\mathrm{O}}}

\usepackage[superscript,biblabel]{cite}

\begin{document}

\mainmatter 

\bibliographystyle{ama}

\title{Static Level Ancestors in Practice}

\author{Matthew Mabrey\inst{1} \and Thomas Caputi\inst{1} \and Georgios Papamichail\inst{2}\and Dimitris Papamichail\inst{1}}

\institute{Department of Computer Science,\\
The College of New Jersey,\\
Ewing, NJ, USA\\
\email{papamicd@tcnj.edu}\\
\url{www.tcnj.edu/~papamicd}
\and
Department of Informatics,\\
New York College,\\
Athens, Greece\\
}

\maketitle

\begin{abstract}
Given a rooted tree $T$, the level ancestor problem aims to answer queries of the form $LA(v, d)$,
which identify the level $d$
ancestor of a node $v$ in the tree. Several algorithms of varied complexity have been
proposed for this problem in the literature, including optimal solutions that preprocess the tree $T$ in linear bounded time and
proceed to answer queries in constant time.
Despite its significance and numerous applications,
to date there have been no comparative studies of the performance of these algorithms and few
implementations are widely available. In our experimental study we  implemented and compared several
solutions to the level ancestor problem, including three theoretically optimal algorithms, and examined
their space requirements and time performance in practice.
\end{abstract}


\section{Introduction and background}

The Level Ancestor (LA) problem is a fundamental problem on trees, defined as follows: 
Given a rooted tree $T$ with $n$ nodes, and queries of the form LA($v, d$), where $v$ is a tree node and $d$ a non-negative integer, find the depth $d$ ancestor of node $v$, meaning the 
$d$th vertex on the path from the root to $v$. Naively, such queries can be answered in $\bigO(n)$ without
the use of any auxiliary data structure,
or alternatively one could precompute all possible queries in $\bigO(n^2)$ time and space, allowing for constant time bounded queries.

The level ancestor problem is related to the extensively studied Least Common Ancestor (LCS) problem. Indeed,
Harel and Tarjan \cite{Ha84} use level ancestors as a subroutine of an LCA algorithm. Level ancestors are used
as part of space-efficient ordinal trees \cite{Ge06}, which can be used for the representation of XML documents
that support XPath queries. Level ancestor queries are part of the primitive operations that are supported 
by other compressed data structures \cite{Sa06} and are used in range-aggregate queries in trees \cite{Yu09}.

There are several known optimal solutions to the level ancestor problem with $\bigO(n)$ complexity 
for preprocessing and 
$\bigO(1)$ query time. Dietz \cite{Di91} first published an optimal algorithm for the dynamic
version of the problem, where Berkman and Vishkin \cite{Be94} published an optimal parallel (PRAM) algorithm,
albeit involving unwieldy constants of the order of $2^{2^{28}}$. It is noteworthy that complex algorithms
for the dynamic and parallel versions of the level ancestor problem preceded simpler algorithms for
the static serial version. A substantially simplified algorithm for the dynamic and static variants 
was published by Alstrup 
and Holm \cite{Al00}, and an even simpler optimal algorithm -- due to its progressive construction in stages --
for the static version was given by Bender and Farach-Colton \cite{Be04}. Ben-Amram \cite{Be09} contributed
yet another simple optimal algorithm for the static level ancestor problem in a technical report, along with
an efficient implementation.

The remainder of this paper is organized as follows. In Section 2 we provide definitions and review the 
algorithms involved in this study. In section 3 the implementations of each algorithm are briefly explained. In Section 4 we describe the experiments performed and the results
obtained. Finally, in Section 5 we draw conclusions about the advantages and limitations
of the different algorithms compared.

\section{Definitions, experimental design and implementations}

Following the notation and definitions in \cite{Be04}, the {\em depth} of a node $v$ in tree $T$, denoted
as $depth(v)$, is the number of edges on the shortest path from $v$ to the root. The root itself has depth
$0$. The {\em height} of a node $v$ in tree $T$, denoted $height(v)$, is the number of nodes on the path 
from $v$ to its deepest descendant. Tree leaves have height $1$. The level ancestor of node $v$ at depth $d$, 
denoted as LA($v,d$), is a node $u$, such that $u$ is
an ancestor of $v$ and $depth(u) = d$. If such a node does not exist, then LA($v,d$) is undefined.
The algorithms described in this paper have both preprocessing and query time complexity. An
algorithm that has preprocessing time $f(n)$ and query time $g(n)$ will be denoted as having
complexity $\langle f(n), g(n) \rangle$.

For the purpose of this study we have implemented eight algorithms for the level ancestor problem, each
with distinct time and space complexies. Four of the algorithms are components of the optimal algorithm
of Bender and Farach-Colton, described in \cite{Be04}, and are eventually combined. The eight algorithms 
implemented are listed below, together with their preprocessing and query time bounds:

\begin{enumerate}
\item Table algorithm -- $\langle \bigO(n^2), \bigO(1)\rangle$
\item Jump-Pointers algorithm -- $\langle \bigO(n\log n), \bigO(\log n)\rangle$
\item Ladder algorithm -- $\langle \bigO(n), \bigO(\log n)\rangle$
\item Jump-Ladder algorithm -- $\langle \bigO(n\log n), \bigO(1)\rangle$
\item Macro-Micro-Tree algorithm -- $\langle \bigO(n), \bigO(1)\rangle$
\item Menghani \& Matani algorithm -- $\langle \bigO(n), \bigO(\log n)\rangle$
\item Ben-Amram algorithm -- $\langle \bigO(n), \bigO(1)\rangle$
\item Hagerup algorithm -- $\langle \bigO(n), \bigO(1)\rangle$
\end{enumerate}

Each of these algorithms have their own efficiency/simplicity trends. 

\section{Level Ancestor Implementations}

In all our implementations the tree is stored in an array, based on the {\em Depth First Search} (DFS) numbering. 
This representation, which is equivalent to a pointer-based tree structure for the static Level Ancestor problem, helps avoid 
a level of indirection when locating a node based on its DFS number. In addition, a DFS procedure only has to visit the table contents in order. This representation can be extended for dynamic approaches, albeit with loss of some of the advantages mentioned. The 
space utilized by this representation is optimal. Every node stores its parent, left and right child pointers, as well as a 
field for the depth of the node, which is required by all algorithms we  discuss, except for the table algorithm.

\subsection{The Table Algorithm}

The table algorithm for the level ancestor problem is the naive solution that preprocesses
and stores all distinct $\bigO(n^2)$ queries. The simplest of all algorithms presented, it is implemented 
with dynamic programming, given the Euler tour of the tree, which we will call the 
{\em tree signature}. A 2-dimensional table holds the level 
ancestors of all nodes. The ancestor list of each node can be progressively generated from the parent node ancestor list, 
when traversing the tree in the DFS order.

This is an optimal query time algorithm, able to answer level ancestor queries by performing only two 
memory references per query, an attribute which makes this algorithm superior to any other solution when 
space is not an issue. The Table algorithm constitutes a compelling solution when dealing with balanced trees or
at least trees with logarithmic bounded expected node depth, and the programming language of choice
supports "ragged" arrays (or arrays of arrays), as is the case with C and Java. Preprocessing time is roughly 
proportional to the space used to store the array.

\subsection{The Jump-Pointer Algorithm}

The jump-pointer algorithm associates every node in the tree with a list of pointers to its ancestor nodes,
that allow to ``jump'' up the tree by powers of $2$. Specifically, for every node $v$ in the tree,
there exists a list associated with $v$ that contains pointers to all $2^i$th ancestors of $v$, for
$0 \le i \le 2^{\lfloor \log depth(v)\rfloor}$.

Using dynamic programming, the pointer lists can be generated in $\bigO(n \log n)$ time by an Euler
tour of the tree.
Queries can be processed in $\bigO(\log n)$ time by following jump pointers up the tree, covering at least
half of the remaining distance to the desired ancestor with each jump. 

\subsection{The Ladder Algorithm}

This algorithm starts with a longest path decomposition of the tree $T$, 
into non-disjoint paths called {\em ladders}. It proceeds by extending the 
ladders towards the root, up to twice their original size. 
This latter action creates the property that every node of {\em height} $h$ belongs to a ladder that includes
its ancestor of height at least $2h$, or the root of the tree, allowing for queries in $\bigO(\log n)$ time.

The ladder algorithm can be implemented by a bottom-up scan of the tree, in which the heights 
of the nodes are calculated. Ladders can be constructed recursively, starting from the deepest node and queuing 
nodes encountered while traversing the current ladder, which are subsequently processed. Careful implementation
leads to a linear time and space algorithm, which, as will be shown in the experimental section below, leads
to significant space savings, while still being competitive in the time domain.

\subsection{The Jump-Ladder Algorithm}

The jump-ladder algorithm combines the preprocessing benefits of the jump-pointer and ladder algorithms 
to achieve constant time queries. Since the jump-pointer algorithm makes exponentially decreasing hops up 
the tree and the ladder algorithm makes exponentially increasing hops up the tree, following a single jump
pointer and then climbing a single ladder leads to the desired level ancestor in two steps.

The preprocessing stage combines the jump-pointer and ladder algorithm preprocessing procedures, resulting
in $\bigO(n \log n)$ complexity in time and space, where the actual space utilized is the sum of 
the space required by both component algorithms. 

\subsection{The Macro-Micro-Tree Algorithm}

The Macro-Micro-Tree algorithm combines and extends the jump-pointer and ladder algorithms. Since
ladders can be preprocessed in linear time, they are used without modifications. Jump pointers
though are not assigned to all nodes -- since the preprocessing would then require $\bigO(n \log n)$ space
and time -- but only to a specific set of nodes, which are called {\em jump nodes}. Ancestors of jump
nodes can use their jump pointers, and are called {\em macro nodes}, forming a connected subtree
of $T$ called the {\em macrotree}. Descendants of jump nodes form disconnected {\em microtrees}.
By applying the standard data structural technique in \cite{Ga83} and limiting the size of 
the microtrees to $\lceil \log n/4 \rceil$, the number of jump nodes is bounded by 
$n / \log n$ and the total number of jump pointers becomes $\bigO(n)$. Due to their limited size,
it has been shown that microtrees adopt a limited number of shapes, $\bigO(\sqrt{n})$, which can
be precomputed and stored using the table algorithm, also in linear bounded time.

The Macro-Micro algorithm has optimal time complexity, but also significant implementation complexity, 
with quite a few details and subtleties on top of the other four implementations. In addition to the node depth and height 
pre-calculations, the {\em weight} of each node, as in the size of the subtree rooted at a node, has to be calculated
in order to characterize Micro and Macro nodes. Overall there are four distinct types of nodes, 
the Micro, the Macro, the Micro-Root (roots of the Micro trees) and the Jump nodes, the latter been 
associated with jump pointers, which can be utilized by their Macro ancestors. Extra space is required to store 
auxiliary pointers for each node, in addition to data structures that hold the tables for the Micro trees.

To locate appropriate entries in the tables of the microtrees, the microtree signatures are
enumerated and stored, being indexed when probed by level ancestor queries. The queries themselves have
multi-case functionality, using a different approach to locate the level ancestor of a node based on
its type.

\subsection{Menghani \& Matani algorithm}

One of the simplest algorithms for the static level ancestor problem was developed by Gaurav Menghani and Dhruv Matani \cite{Me19}. The algorithm performs a pre-order traversal of a rooted tree to assign to each node a unique label in the order that it is visited, with the root starting as zero. Node labels are unique (no duplicates) and values range from $0$ to $(n - 1)$, for a tree with $n$ nodes. The algorithm takes advantage of a property of this tree that for any given node $v$, all descendants of $v$ have label values greater than that of $v$ and smaller than the label of $v$'s closest right-sibling node. This property allows us to find a depth $d$ ancestor of any given node $v$ by locating the largest label value node at depth $d$ less than $v$'s label value. In practice, this is accomplished using a two-dimensional array for every depth $d$ containing every node at that depth in ascending label value order. To answer  level-ancestor query LA($v$, $d$), a modified binary search is performed on the array for depth $d$ that uses $v - 1$ as the goal value and keeps track of the maximum value less then the goal value found. If the goal value is found, then the associated node with that label is the answer to the level ancestor query, otherwise it is the node associated with the maximum value found. This results in a sub-optimal query time of $\bigO(\log n)$, though it can be improved to $\bigO(\log \sqrt{n})$ for sufficiently large arrays using a faster binary search technique.

\subsection{Ben-Amram's Find-Smaller algorithm}

Another relatively simple optimal algorithm for the static level ancestor problem was presented by Ben-Amram
\cite{Be09}, which is also utilizes the microset technique \cite{Ha84,Ga83}. His algorithm uses the
Euler tour representation of a rooted tree and is a simplification of Berkman and Vishkin's PRAM algorithm
\cite{Be94}. It reduces the level ancestor problem to the {\em Find-Smaller} problem, the goal of which
is, given an array $A$, an index $v$ and an integer $d$, to find the minimal $u > v$ such that $A_u \le d$.
Auxiliary array data structures are created to support efficient queries and the microset technique
is used to sparsify these data structures.

Ben Amram's algorithm was implemented in C by Victor Buchnik and can be obtained by contacting the authors. We used
an obtained implementation to compare to our own algorithm implementations described in the other subsections. We will refer to Ben-Amram's algorithm as the {\em Find-Smaller} algorithm.

\subsection{Torben Hagerup's Find-Larger algorithm}

The Find-Larger algorithm created by Torben Hagerup \cite{hagerup2020simpler} is a variation of the previous algorithm which reduces the static level ancestor problem to the Find-Smaller problem. The algorithm inverts the values of the Find-Smaller array $A$ such that the level ancestor problem reduces to finding the minimal $u > v$ such that $A_u \ge d$. The algorithm then constructs a series of data structures from the Find-Larger array which can be represented as a two-dimensional line graph of peaks and valleys, with the root being the tallest peak, and the greatest depth nodes being the lowest valleys. Ladder data structures are created for each index of the Find-Larger array and contain answers for level ancestor queries, with the deepest valleys containing the largest ladders, in some cases answering queries all the way up to the root node. The algorithm then answers a level ancestor query for node $v$ by consulting $v$'s ladder, and if the answer is not contained within, jumping to $v$'s deepest reachable valley and using it’s ladder structure to find the answer.

\subsection{Tree representation and random generation}

All level ancestor algorithms work on rooted trees, which consist part of the input of the algorithm.
To represent rooted trees, we decided to use the Euler representation, a unique mapping of a tree to an
array which is compact and is derived from an Euler traversal of the tree, commonly referred to
as {\em depth first search} (DFS). For our implementations we selected to use a binary representation
of the Euler traversal, where `1' represents a forward downward traversal of an edge towards a descendant
node, and `0' represents an upward traversal of an edge toward an ancestor node. A rooted tree with $n$
nodes can therefore be represented by $2n-2$ binary digits, which we call the {\em signature} of
the tree. It should be noted that, for ordinal binary trees of large size, this is the most 
compact representation.
For the purpose of our study, we have used an ASCII representation of the binary digits of the Euler
traversal, in order to facilitate visual inspection and debugging. Given a rooted tree signature, it is 
trivial to construct a pointer linked or table based tree structure in linear time on the size of the tree.

To generate random trees for our experiments, we use the split subtree method \cite{De96}, where,
in order to create a tree
with $n$ nodes, we generate a real-valued random variable $x$ in the unit interval $(0,1)$, assign 
$\lfloor xn \rfloor$ nodes to the left subtree, one node to the root, and the rest to the right subtree.
The process continues recursively for each subtree. This method generates trees equivalent to ones 
generated by the
random binary search tree process by using random permutations \cite{Mc12}, where any node has an 
equal probability to be chosen as root.

This method has several advantages when constructing random trees. It can be used in a straightforward
manner to construct 
the Euler traversal sequence (signature) of a random tree with $n$ nodes. First we allocate a table with $2n-2$ 
positions and select a random number $k = \lfloor xn \rfloor$ between $1$ and $n$. We can then insert
`1' in positions $0$ and $2k$, `0' in positions $2k-1$ and $2n-3$, and recursively process
the two subtrees and corresponding subarrays (or single subtree/subarray when $k=n$). The expected average 
depth of a
node of a random tree created with the split subtree method is $\bigO(\log n)$ \cite{Al96}, 
while the expected depth of the tree is also bounded logarithmically \cite{De86}. These observations
have been validated in the experiments described in the next section. The split subtree method also is easily
modifiable to accommodate skewed unbalanced trees, by varying the range of values that the random variable takes.
Using that property we have generated skewed trees with higher average node depth, as described in
the next section.

\section{Algorithm Evaluation}

All eight algorithms described in the previous section were implemented in the programming language C, using standard libraries. All programs were compiled with gcc version 7.2.0, on the CentOS Linux 7.9.2009 operating system, running on an AMD EPYC 7702P Zen2 CPU (incorporates logic fabricated TSMC 7nm process and I/O fabricated on GlobalFoundries 14nm process) at 2 GHz, on a single core and thread. Each core has 32KB of L1 cache, 512KB of L2 cache, and shares 8MB of L3 cache with the other cores. The machine used has 512GB of main memory and test cases have been limited in size in order for program instructions and data to reside in main memory. All programs were compiled with the default `-O0' optimization level set.

\begin{figure}
\centering
\subfigure[]{
\includegraphics[height=5.5cm,angle=-90]{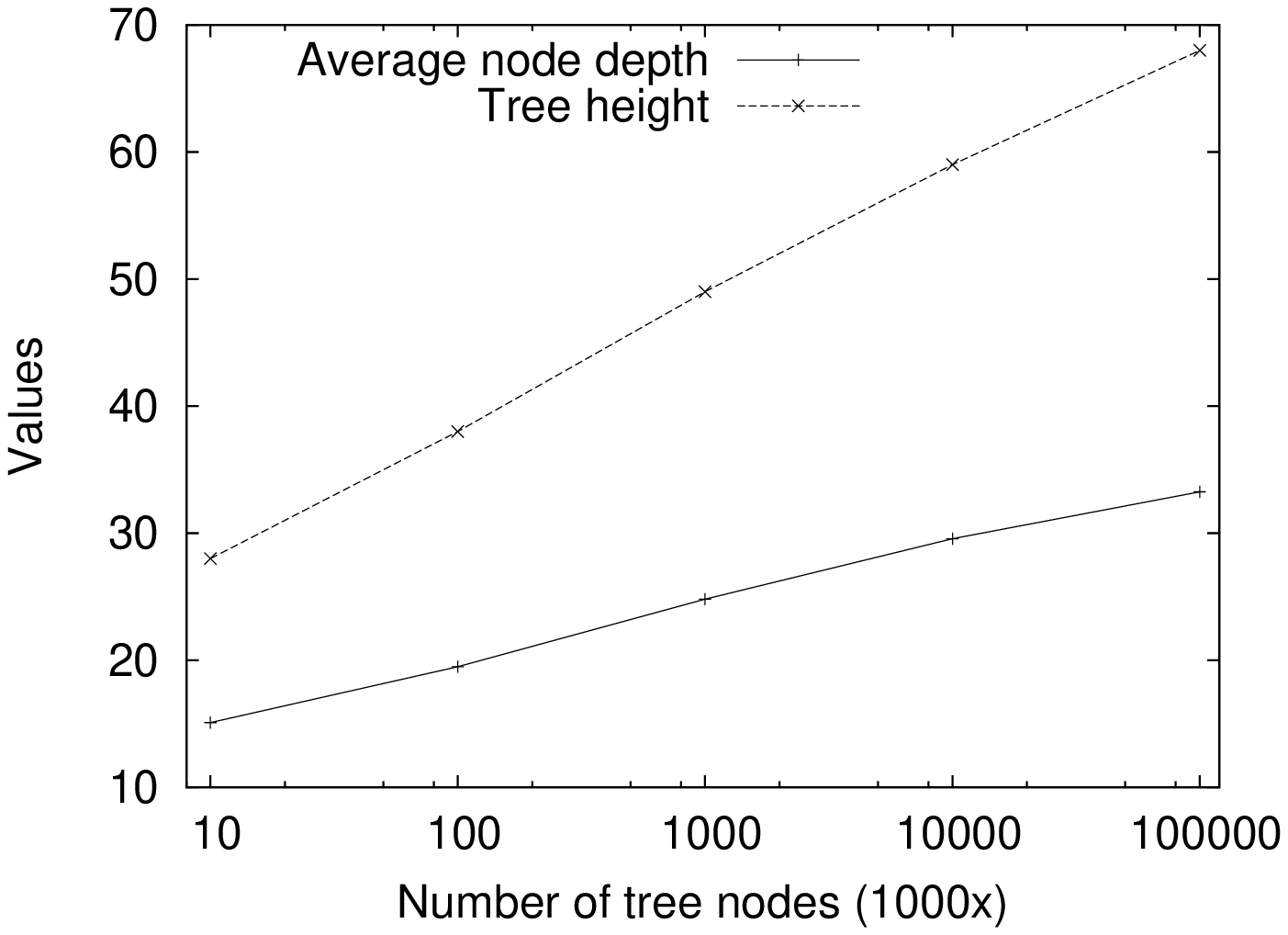}
}
\subfigure[]{
\includegraphics[height=5.5cm,angle=-90]{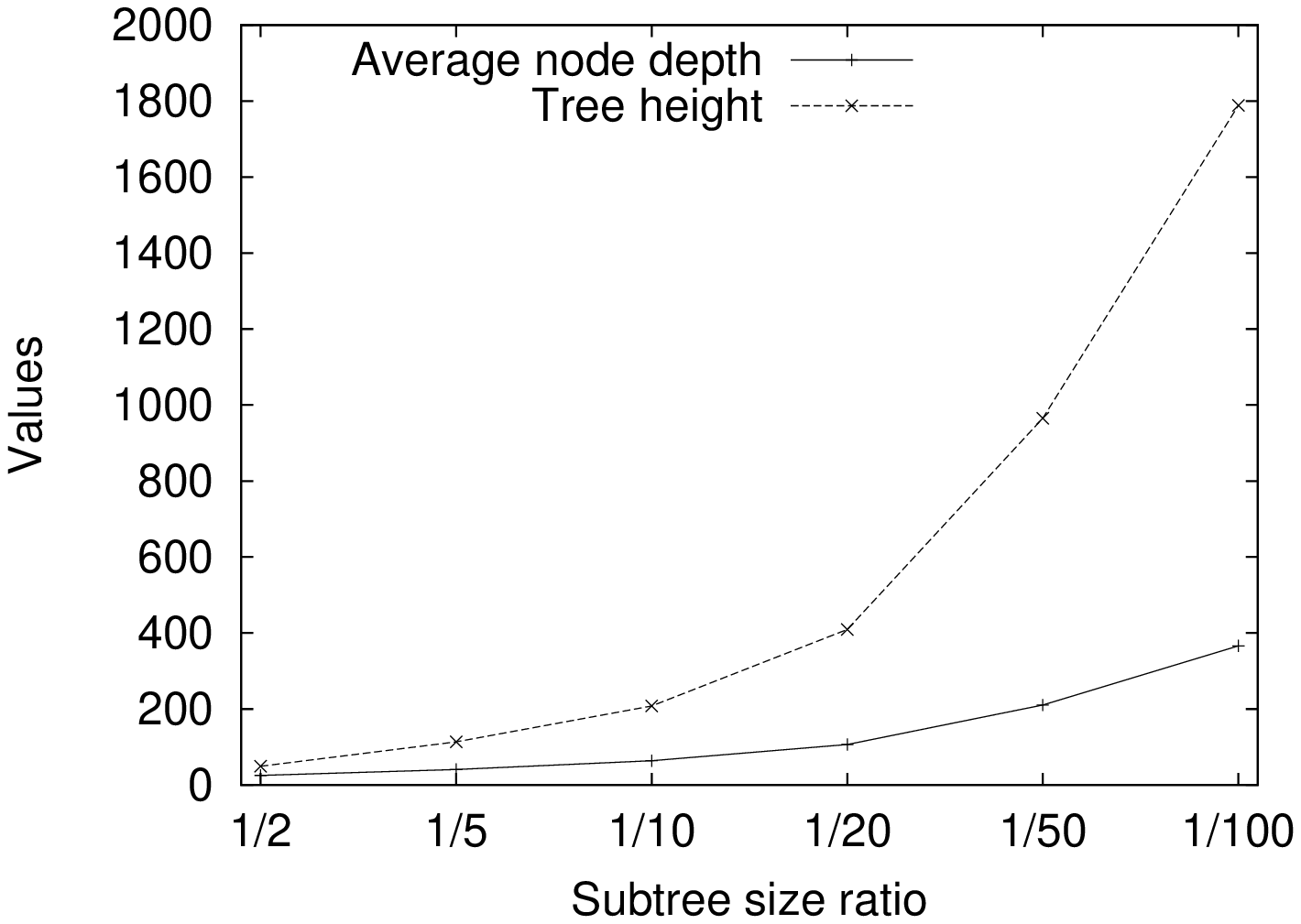}
}
\caption{Tree properties. (a) Randomly generated split trees. (b) Randomly generated split trees with bounded
subtree size ratio.}
\label{fig:tree_stats}
\end{figure}

Randomly generated trees were varied in size between 100K and 1B nodes, 
with sizes increasing by a factor of 10 for each step. For each size step, 5 random trees were generated and the preprocessing and query time results were averaged between 5 runs of each algorithm on each random tree. The number of random level ancestor queries generated and run for each experiment were kept constant at 100M, which we deemed sufficient for experiments to run long enough not to be influenced by process swapping, interrupts, and other operating system events. We decided not to vary the number of queries for each experiment, since query execution commences after static structures are in place following their preprocessing, and we expect the execution time to be a linear function of the number of queries. Tree depth and average node depth for the random trees generated are depicted in Figure \ref{fig:tree_stats}(a). Each algorithm’s preprocessing and query time is calculated using the clock\_gettime() function included in the standard C library time.h header file. We timed algorithms using both elapsed real-time and CPU clock time and, since we had exclusive use of the CPU, we determined there was a negligible difference between the two and ultimately tracked the elapsed real-time for the experiments. 

Applications of the level ancestor problems quite often involve trees that are not balanced and
may have properties, such as depth, that are not well represented by the expected evenly distributed nodes of random
trees described above. To better examine the behavior of the algorithms when run on unbalanced trees, we created
a set of skewed rooted trees with 1M nodes each, varying the size ratio of the subtrees at each node. 
Using the same split subtree method, we limited the range of the random variable to a fraction of the 
unit interval. The properties of these skewed trees generated for varying ratios are shown in Figure
\ref{fig:tree_stats}(b).

\subsection{Preprocessing}

Preprocessing involves the time required to build necessary auxiliary data structures
that support each algorithm, and includes the time it takes to load the tree in main memory. 
Experimental preprocessing times for all eight algorithms and varying
tree sizes, for the randomly generated trees with expected logarithmic height, are shown 
in Table \ref{table:1}.

\begin{table}
\caption{Preprocessing times (seconds) for evenly distributed randomly generated trees}
\begin{center}
\begin{tabular}{|r|r|r|r|r|r|r|r|r|}
\hline
\multicolumn{1}{|c|}{\bf Nodes}&\multicolumn{1}{|c|}{\bf Table}&\multicolumn{1}{|c|}{\bf Jump-}&\multicolumn{1}{|c|}{\bf Ladder}&\multicolumn{1}{|c|}{\bf Jump-}&\multicolumn{1}{|c|}{\bf Macro-}&\multicolumn{1}{|c|}{\bf Menghani}&\multicolumn{1}{|c|}{\bf Find-}&\multicolumn{1}{|c|}{\bf Find-}\\
\multicolumn{1}{|c|}{\bf (1000x)}&&\multicolumn{1}{|c|}{\bf Pointer}&&\multicolumn{1}{|c|}{\bf Ladder}&\multicolumn{1}{|c|}{\bf Micro-Tree}&\multicolumn{1}{|c|}{\bf \& Matani}&\multicolumn{1}{|c|}{\bf Smaller}&\multicolumn{1}{|c|}{\bf Larger}\\
\hline\rule{0pt}{2pt}
100   &   0.017   &   0.011   &   0.011   &   0.016   &   0.015   &   0.006   &   0.022   &   0.014  \\
1000   &   0.140  &   0.082   &   0.097   &   0.143  &   0.140   &   0.044   &   0.169   &   0.106  \\ 
10000   &   1.500   &   0.778   &   0.946   &   1.439   &   1.343  &   0.408   &   1.641   &   1.008  \\ 
100000   &   16.560   &   7.911  &   9.339 &   14.422  &   13.201   &   4.000  &   16.374   &   10.091  \\  
1000000   &   183.366   &   80.510   &   93.745   &   145.759  &   131.124  &   40.259   &   163.571   &   101.523  \\
\hline
\end{tabular}
\end{center}
\label{table:1}
\end{table}

All algorithms can preprocess trees of size up to 100M nodes in less than 17 seconds, with the Menghani \& Matani
algorithm outperforming the rest (e.g. taking about 4 seconds to preprocess 100M nodes vs about 8 seconds for the next best). When tree sizes reach 1B nodes the differing preprocessing bounds become even more apparent as several algorithms preprocessing times take more than 100 seconds, with the Table algorithm being the worst at around 180 seconds for preprocessing. The Menghani \& Matani algorithm continues to perform the best in terms of preprocessing speed as it takes about 40 seconds to preprocess a tree with 1B nodes. While the differences between algorithms seem significant, preprocessing is a one-time cost for static trees; no algorithm seems to have a clear disadvantage based on preprocessing time alone when tree size remains at or below 1B nodes. In the case of skewed trees,
all algorithms except the Table algorithm perform equally well (data not shown). The Table algorithm experiences
preprocessing time increases proportional to its space utilization, as analyzed in Table \ref{table:1}.

\subsection{Space Requirements}

Space is probably the most critical factor in selecting an appropriate level ancestor algorithm.
Queries can be performed in constant or near-constant time and preprocessing is expedient even for very large trees.
Since most solutions will be satisfactory when processing small trees in modern workstations, it 
is large trees and the extra storage space requirements which will influence most significantly the selection
of one solution over another. The space utilization of the eight algorithm implementations after preprocessing is concluded
is presented in Tables \ref{table:2} and \ref{table:3}. This data was collected using the maximum resident set size statistic value displayed by the Linux $/usr/bin/time$ command using the $-v$ GNU option.

\begin{table}
\caption{Space usage (MB) of level ancestor algorithms for random trees with log bounded expected depth}
\begin{center}
\begin{tabular}{|r|r|r|r|r|r|r|r|r|}
\hline
\multicolumn{1}{|c|}{\bf Nodes}&\multicolumn{1}{|c|}{\bf Table}&\multicolumn{1}{|c|}{\bf Jump-}&\multicolumn{1}{|c|}{\bf Ladder}&\multicolumn{1}{|c|}{\bf Jump-}&\multicolumn{1}{|c|}{\bf Macro-}&\multicolumn{1}{|c|}{\bf Menghani}&\multicolumn{1}{|c|}{\bf Find-}&\multicolumn{1}{|c|}{\bf Find-}\\
\multicolumn{1}{|c|}{\bf (1000x)}&&\multicolumn{1}{|c|}{\bf Pointer}&&\multicolumn{1}{|c|}{\bf Ladder}&\multicolumn{1}{|c|}{\bf Micro-Tree}&\multicolumn{1}{|c|}{\bf \& Matani}&\multicolumn{1}{|c|}{\bf Smaller}&\multicolumn{1}{|c|}{\bf Larger}\\
\hline\rule{0pt}{2pt}
100   & 13  & 6  & 5  & 9  & 7  & {\bf 3}  & 6 & 15 \\
1000   & 136 & 55  & 48  & 87  & 61  & {\bf 32}  & 55 & 146 \\
10000   & 1569  & 548  & 461  & 852  & 592  & {\bf 210}  & 531  & 1457  \\
100000   & 17495  & 5470  & 4586  & 8494  & 5840  & {\bf 1966}  & 5132 & 14594  \\
1000000  & 199845 & 53697  & 45860  & 84934  & 57815  & {\bf 19547}  & 50198 & 146175 \\
\hline
\end{tabular}
\end{center}
\label{table:2}
\end{table}

The Menghani \& Matani algorithm performs the best when it comes to space utilization. It consistently pulls ahead of the more complex constant query time algorithms, an effect we primarily attribute due to its particularly simple data structures.

Notably, all algorithms have a linear space utilization dependence on the size of the tree, as
demonstrated in the results shown in Table \ref{table:2}, with the exception for the Table algorithm, which
shows $\bigO(n\log h)$ space utilization, where $h$ is the average node depth of the tree. The last
test case set with trees of 1B nodes forced the Table algorithm to use approximately
200GB of memory, which significantly affected the preprocessing time of the algorithm as well.

\begin{table}
\caption{Space usage (MB) of level ancestor algorithms using skewed random trees}
\begin{center}
\begin{tabular}{|r|r|r|r|r|r|r|r|r|}
\hline
\multicolumn{1}{|c|}{\bf Ratio}&\multicolumn{1}{|c|}{\bf Table}&\multicolumn{1}{|c|}{\bf Jump-}&\multicolumn{1}{|c|}{\bf Ladder}&\multicolumn{1}{|c|}{\bf Jump-}&\multicolumn{1}{|c|}{\bf Macro-}&\multicolumn{1}{|c|}{\bf Menghani}&\multicolumn{1}{|c|}{\bf Find-}&\multicolumn{1}{|c|}{\bf Find-}\\
\multicolumn{1}{|c|}{\bf }&&\multicolumn{1}{|c|}{\bf Pointer}&&\multicolumn{1}{|c|}{\bf Ladder}&\multicolumn{1}{|c|}{\bf Micro-Tree}&\multicolumn{1}{|c|}{\bf \& Matani}&\multicolumn{1}{|c|}{\bf Smaller}&\multicolumn{1}{|c|}{\bf Larger}\\
\hline\rule{0pt}{2pt}
1/2   & 137  & 55  & 45  & 84  & 59  & \bf{32}  & 55  & 147\\
1/5   & 198  & 56  & 42  & 82  & 58  & \bf{32}  & 56  & 147\\
1/10  & 295  & 63  & 41  & 87  & 57  & \bf{32}  & 56  & 147\\
1/20  & 448  & 68  & 40  & 92  & 57  & \bf{32}  & 57  & 147\\
1/50  & 868  & 70  & 38  & 92  & 51  & \bf{32}  & 58  & 148\\
1/100 & 1479  & 71  & 39  & 94  & 54  & \bf{32}  & 58 & 148\\
\hline
\end{tabular}
\end{center}
\label{table:3}
\end{table}

In the case of unbalanced trees, as observed in Figure \ref{table:3}, the Menghani \& Matani and Ladder algorithms exhibit the lowest space utilization, followed by the Macro-Micro-Tree and Hagerup algorithms. An interesting trend is noticeable in
the space utilization of the Ladder algorithm, which seems to slightly improve with increased imbalance.
This is not surprising, since unbalanced trees are expected to have fewer ladders as a result
of the path decomposition, and fewer additional nodes on average when they are extended. Additionally, the Mengani \& Matani algorithm is not affected by increased imbalance. This can be attributed to the simple data structure used by the algorithm which generally stores a single integer for each node and an integer pointer for each depth value, quantities that do not significantly change with increased imbalance.

\begin{figure}
\centering
\includegraphics[height=10cm,angle=-90]{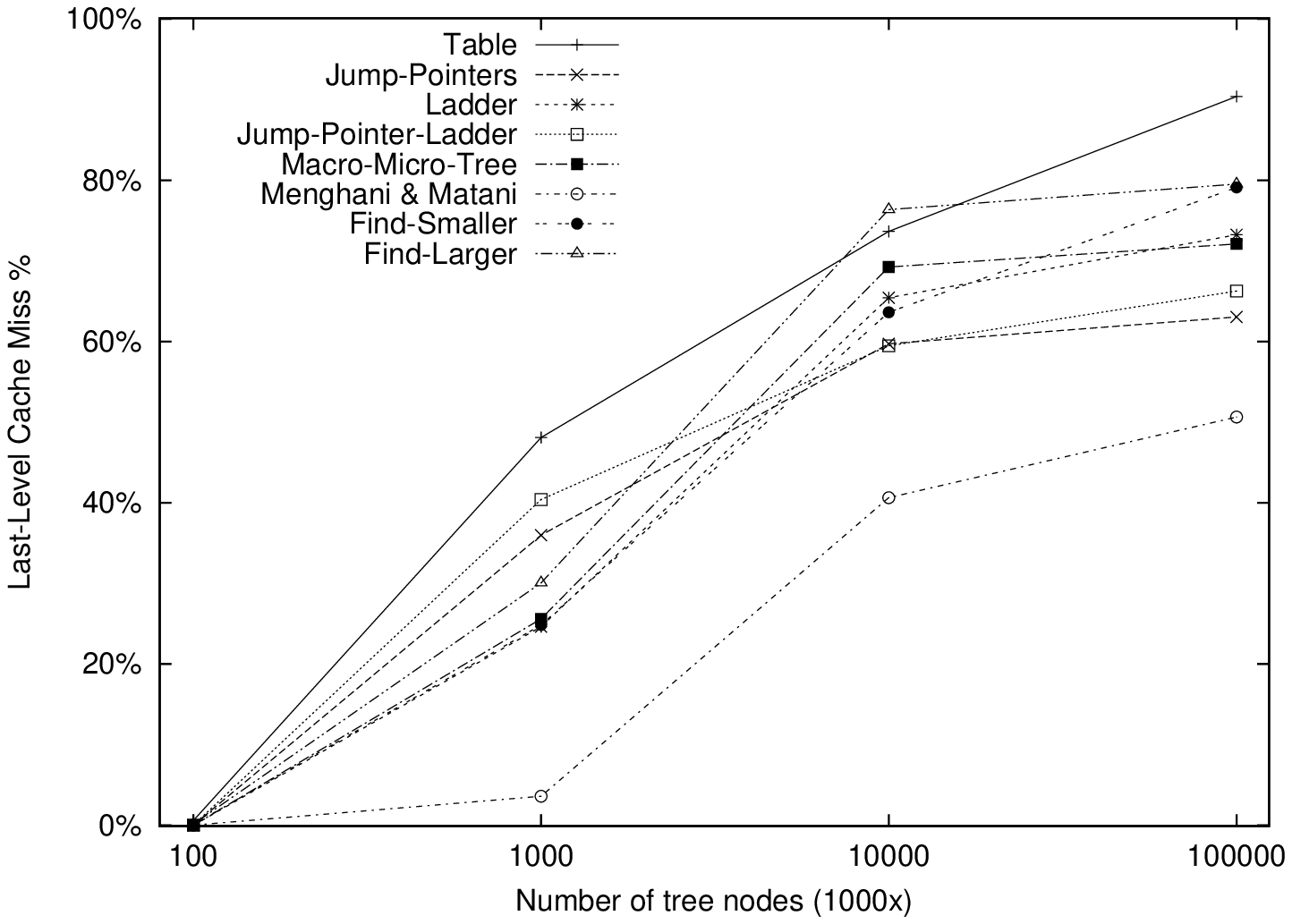}
\caption{Percentage of last-level cache (LLC) references that result in a cache miss}
\label{fig:cache_util}
\end{figure}

Locality of reference of an algorithm is another significant factor when selecting a specific implementation, as algorithms which repeatedly access the same data in answering level ancestor queries are expected to suffer less from cache misses. The locality of reference of each examined algorithm is partially illustrated in Figure \ref{fig:cache_util}\footnotemark[1], which shows how often each algorithm suffers a last-level cache (LLC) miss and reads data from main memory. There seems to be a correlation between algorithm query time and LLC misses when compared to query times in Figure \ref{fig:queries}. Even though the number of queries is fixed for each experiment, query times continue to grow as LLC miss percentage approaches 100\%. Additionally, the Menghani \& Matani algorithm's query time isn't nearly as impacted as the query time of all of the other algorithms when tree size steps from 100k to 1 million nodes. This can possibly be explained by the algorithm's superior locality of reference and having 96.4\% of all cache references result in a cache hit at a tree size of 1 million nodes. 

\footnotetext[1]{ Note: Figure \ref{fig:cache_util} shows data for trees having up to 100 million nodes; the AMD-based machine used for the rest of the experiments did not provide access to hardware event data such as cache references and misses, so this experiment was conducted on an Intel-based machine. As this latter machine had smaller main memory, experimental tree size was limited in order to maintain trees in memory. Otherwise, the behavior of the examined algorithms on the machines with different processor architectures was comparable.}

\subsection{Query time performance}

\begin{figure}
\centering
\includegraphics[height=12cm,angle=-90]{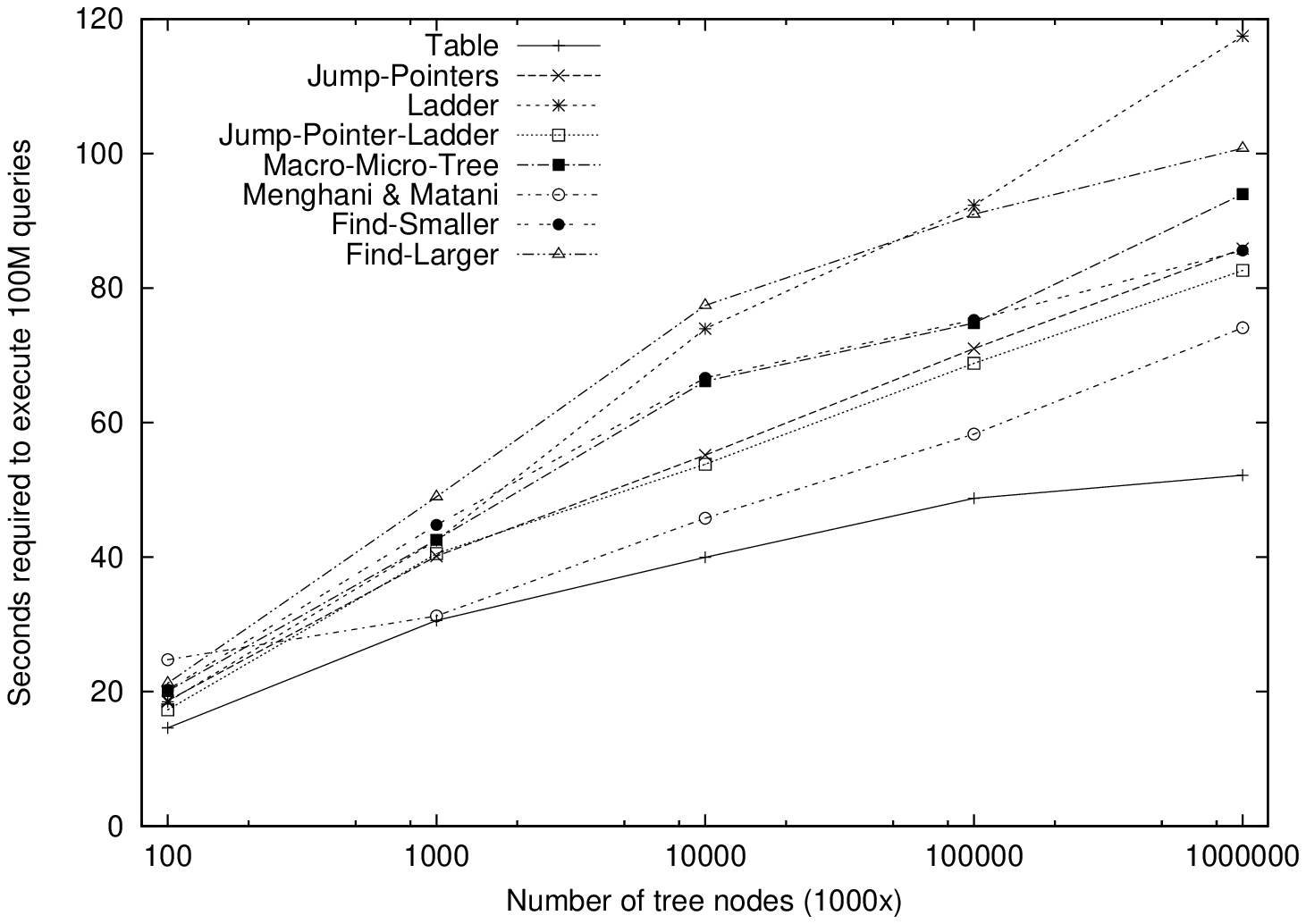}
\caption{Time to run 100M queries on randomly generated trees}
\label{fig:queries}
\end{figure}

Level ancestor queries are often called as subroutines in other algorithms. As such, it is critical
that they are expediently executed. In Figure \ref{fig:queries} we can observe that the Table algorithm
outperforms all others in query execution time as long as its data structures can be stored in main memory in their entirety, which in our experiments was possible due to the available 512GB of main memory. In scenarios where the table algorithm cannot fit its data structures in main memory, a significant number of page faults will drastically affect the query time of the algorithm. All algorithms, including the query-optimal ones, experience increased query execution times as a function of tree size, which may be explained by increased cache misses, once their data structures exceed the size of the different levels of
cache of the processor. 
Surprisingly, the Menghani \& Matani algorithm with its sub-optimal query time outperforms all but the Table algorithm once the tree size reaches 1M nodes. One possible explanation would involve the small, frequently accessed data structures, leading to higher cache memory utilization.

\begin{figure}
\centering
\includegraphics[height=12cm,angle=-90]{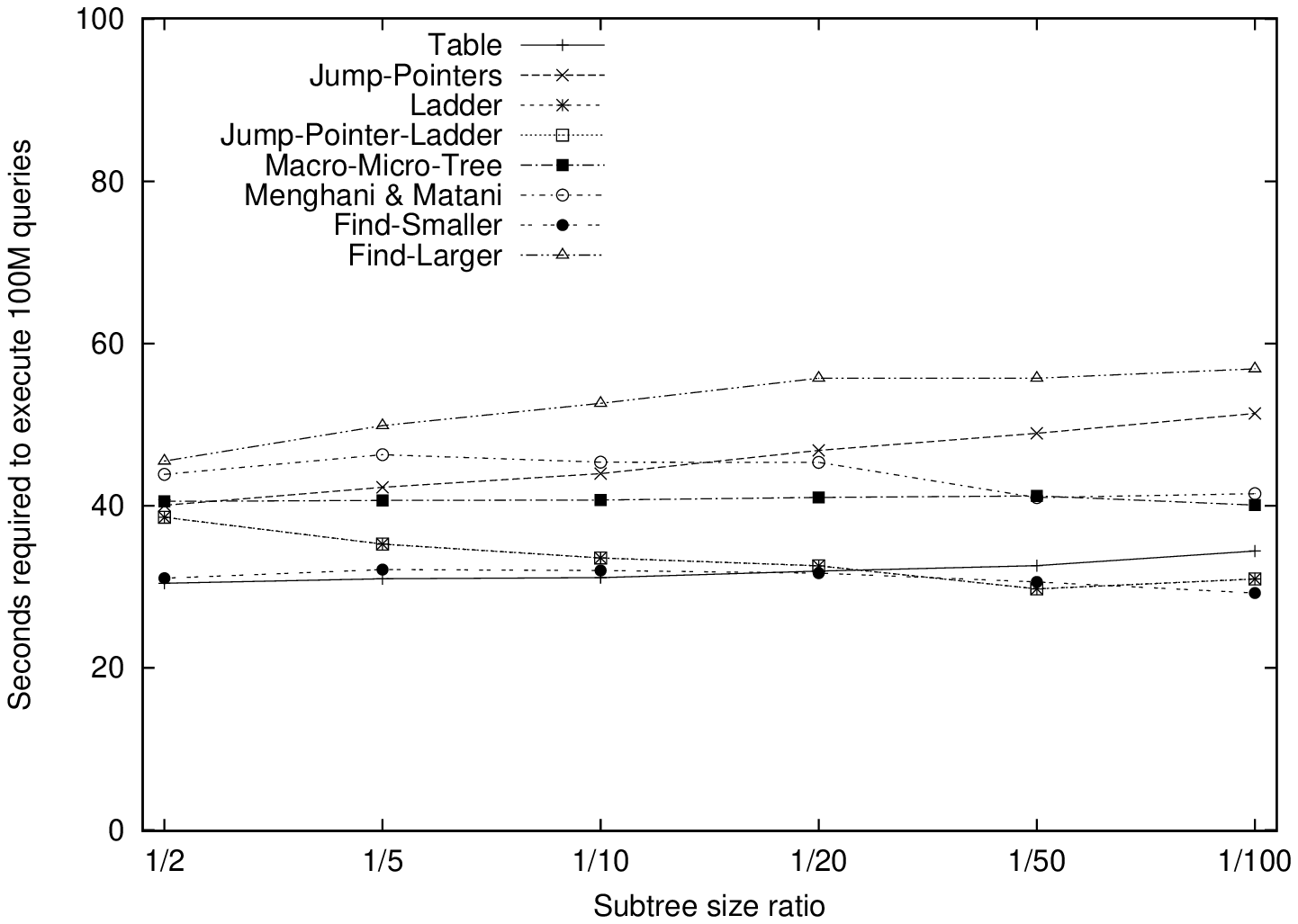}
\caption{Time to run 100M queries on unbalanced randomly generated trees with 1M nodes. The x axis
represents the maximum ratio between subtree sizes of node children.}
\label{fig:queries_skewed}
\end{figure}

On unbalanced trees, the Ladder and Menghani \& Matani algorithms perform particularly well, with query times
that improve as imbalance increases. Once again, the Ladder algorithm is expected to form fewer ladders as a result
of tree imbalance, which can lead to a reduced expected number of hops between ladders when executing
queries. For the Menghani \& Matani algorithm, the smaller arrays at each depth in an imbalanced tree can lead to faster binary search resolutions, which in turn speed up queries for the algorithm. The performance of all other algorithms seem to remain relatively stable with increased
imbalance, except for the Jump-Pointer algorithm, which seems to require an increasing number of 
jumps for nodes to reach their ancestors, as the average node depth of the tree increases.

\section{Conclusions}

Each algorithm presented, implemented, and tested in this paper has distinct merits and shortcomings. 
No one algorithm has a clear advantage over the others in every aspect, although the Menghani \& Matani algorithm outperforms most in several areas. The Table algorithm outperforms -- 
as expected -- all other solutions in query execution time, if one can afford its worst case quadratic 
space requirement, which can be significant even when the input tree has logarithmically bounded depth. 
Space utilization, which is possibly the most critical component influencing algorithm choice,
seems to correlate with the simplicity of algorithms with data structures that use linear bounded space, with the Menghani \& Matani, Jump-Pointer-Ladder, and Find-Smaller algorithms outperforming all others, and
the first two demonstrating increased benefits when dealing with unbalanced trees. 

Some of the simpler non-optimal algorithms, such as the Menghani \& Matani, lead to simple efficient implementations, that are easier to comprehend, maintain, optimize and extend. 
On the other hand, with ever increasing problem sizes and machine capabilities, the optimal algorithms, especially the Find-Smaller, may be the solutions of choice
once tree size sufficiently large, but certainly much larger than 1B nodes. Overall, the Menghani \& Matani and Ben-Amram's algorithms seem to be the most versatile solutions, combining excellent query execution times with a 
smaller space footprint and reasonable preprocessing requirements.

The Table, Jump-Pointer, Ladder, Jump-Ladder, Macro-Micro-Tree, Menghani \& Matani, and Hagerup algorithm implementations can be downloaded at \url{www.tcnj.edu/~papamicd/level_ancestor}.
The Find-Smaller algorithm implementation can be obtained from its authors \cite{Be09}.

\section*{Acknowledgments.} The authors would like to acknowledge the assistance of Shawn Sivy who graciously provided critical hardware to perform our experiments. The authors also acknowledge use of the ELSA high performance computing cluster at The College of New Jersey for all computational experiments performed. This cluster is funded by the NSF grant OAC-1828163.


\bibliography{la}

\begin{thebibliography}{10}

\bibitem{Al96}
David Aldous.
\newblock Probability distributions on cladograms.
\newblock In {\em In Random Discrete Structures}, pages 1--18. Springer, 1996.

\bibitem{Al00}
Stephen Alstrup and Jacob Holm.
\newblock Improved algorithms for finding level ancestors in dynamic trees.
\newblock In {\em Automata, Languages and Programming, 27th International
  Colloquium, ICALP 2000, number 1853 in LNCS}, pages 73--84. Springer Verlag,
  2000.

\bibitem{Be09}
Amir~M. Ben-Amram.
\newblock The euler path to static level-ancestors.
\newblock {\em CoRR}, abs/0909.1030, 2009.

\bibitem{Be04}
Michael~A. Bender and Mart\'{\i}n Farach-Colton.
\newblock The level ancestor problem simplified.
\newblock {\em Theor. Comput. Sci.}, 321(1):5--12, June 2004.

\bibitem{Be94}
Omer Berkman and Uzi Vishkin.
\newblock Finding level-ancestors in trees.
\newblock {\em Journal of Computer and System Sciences}, 48(2):214 -- 230,
  1994.

\bibitem{De86}
Luc Devroye.
\newblock A note on the height of binary search trees.
\newblock {\em J. ACM}, 33(3):489--498, May 1986.

\bibitem{De96}
Luc Devroye and Paul Kruszewski.
\newblock The botanical beauty of random binary trees.
\newblock In Franz Brandenburg, editor, {\em Graph Drawing}, volume 1027 of
  {\em Lecture Notes in Computer Science}, pages 166--177. Springer Berlin /
  Heidelberg, 1996.
\newblock 10.1007/BFb0021801.

\bibitem{Di91}
Paul Dietz.
\newblock Finding level-ancestors in dynamic trees.
\newblock In Frank Dehne, Jörg-Rüdiger Sack, and Nicola Santoro, editors,
  {\em Algorithms and Data Structures}, volume 519 of {\em Lecture Notes in
  Computer Science}, pages 32--40. Springer Berlin / Heidelberg, 1991.
\newblock 10.1007/BFb0028247.

\bibitem{Ga83}
Harold~N. Gabow and Robert~Endre Tarjan.
\newblock A linear-time algorithm for a special case of disjoint set union.
\newblock In {\em Proceedings of the fifteenth annual ACM symposium on Theory
  of computing}, STOC '83, pages 246--251, New York, NY, USA, 1983. ACM.

\bibitem{Ge06}
Richard~F. Geary, Rajeev Raman, and Venkatesh Raman.
\newblock Succinct ordinal trees with level-ancestor queries.
\newblock {\em ACM Trans. Algorithms}, 2(4):510--534, October 2006.

\bibitem{Ha84}
Dov Harel and Robert~Endre Tarjan.
\newblock Fast algorithms for finding nearest common ancestors.
\newblock {\em SIAM J. Comput.}, 13(2):338--355, May 1984.

\bibitem{Mc12}
C.C. McGeoch.
\newblock {\em A Guide to Experimental Algorithmics}.
\newblock Cambridge University Press, 2012.

\bibitem{Sa06}
Kunihiko Sadakane and Roberto Grossi.
\newblock Squeezing succinct data structures into entropy bounds.
\newblock In {\em Proceedings of the seventeenth annual ACM-SIAM symposium on
  Discrete algorithm}, SODA '06, pages 1230--1239, New York, NY, USA, 2006.
  ACM.

\bibitem{Yu09}
Hao Yuan and Mikhail~J. Atallah.
\newblock Efficient data structures for range-aggregate queries on trees.
\newblock In {\em Proceedings of the 12th International Conference on Database
  Theory}, ICDT '09, pages 111--120, New York, NY, USA, 2009. ACM.

\end{thebibliography}

\end{document}